\documentclass[conference]{IEEEtran}
\IEEEoverridecommandlockouts

\usepackage{amsmath,amssymb,amsfonts}   %
\usepackage{textcomp}                   %

\usepackage{graphicx}                   %

\usepackage{array}                      %
\newcolumntype{H}{>{\setbox0=\hbox\bgroup}c<{\egroup}@{}} %
\usepackage{booktabs}                   %
\usepackage{siunitx}                    %
  \DeclareSIUnit{\dBm}{dBm}
  \DeclareSIUnit{\dBi}{dBi}
  \DeclareSIUnit{\dBsm}{dBsm}

\usepackage{soul}                       %

\usepackage[backend=biber,
            language=english,
            autolang=other,
            style=ieee]{biblatex}       %
\usepackage[dvipsnames,svgnames,table]{xcolor} %
\usepackage[hidelinks]{hyperref}               %
\usepackage[capitalize]{cleveref}                          %

\usepackage{algorithmic}                %

\usepackage{pgfplots}                   %
  \pgfplotsset{compat=newest}
\usepgfplotslibrary{groupplots,dateplot,polar} %
\usepackage{pgfplotstable}              %
  \pgfplotsset{compat=1.18}              %

\usetikzlibrary{%
  patterns,          %
  shapes.arrows,     %
  spy,               %
  external,          %
  calc,              %
  shadings,          %
  shadows.blur,      %
  decorations.pathreplacing %
}

\usepackage[xindy,nonumberlist]{glossaries} %
  \makenoidxglossaries            %
  
\newacronym{2d}{2D}{two-dimensional}
\newacronym{3d}{3D}{three-dimensional}
\newacronym{3gpp}{3GPP}{3rd generation partnership project}
\newacronym{3msk}{3MSK}{3-level minimum-shift-keying}
\newacronym{4g}{4G}{fourth-generation}
\newacronym{5g}{5G}{fifth-generation}
\newacronym{6dof}{6DoF}{six degrees of freedom}
\newacronym{6g}{6G}{sixth-generation}
\newacronym{abp}{ABP}{authentication by personalisation}
\newacronym{ac}{AC}{alternating current}
\newacronym{aci}{ACI}{adjacent channel interference}
\newacronym{aclr}{ACLR}{adjacent channel leakage ratio}
\newacronym{acpr}{ACPR}{adjacent channel power ratio}
\newacronym{adc}{ADC}{analog-to-digital converter}
\newacronym{adr}{ADR}{adaptive data rate}
\newacronym{aec}{AEC}{aluminum electrolytic capacitor}
\newacronym{aes}{AES}{Advanced Encryption Standard}
\newacronym{afe}{AFE}{analog front end}
\newacronym{ag}{AG}{array gain}
\newacronym{agc}{AGC}{automatic gain controller}
\newacronym{agps}{A-GPS}{Assisted Global Positioning System} %
\newacronym{ai}{AI}{artificial intelligence}
\newacronym{alk}{ALK}{Alkaline}
\newacronym{am}{AM}{amplitude modulation}
\newacronym{amam}{AM/AM}{amplitude modulation to amplitude modulation}
\newacronym{amc}{AMC}{automatic modulation classification}
\newacronym{amp}{AMP}{approximate message passing}
\newacronym{ampm}{AM/PM}{amplitude modulation to phase modulation}
\newacronym{aoa}{AOA}{angle-of-arrival}
\newacronym{aod}{AOD}{angle-of-departure}
\newacronym{ap}{AP}{access point}
\newacronym{api}{API}{application program interface}
\newacronym{apt}{APT}{acoustic power transfer}
\newacronym{apu}{APU}{access point unit}
\newacronym{ar}{AR}{augmented reality}
\newacronym{arima}{ARIMA}{Auto-Regressive Integrated Moving Average}
\newacronym{arp}{ARP}{Antenna Reference Point}
\newacronym{asic}{ASIC}{application specific integrated circuit}
\newacronym{ask}{ASK}{amplitude-shift keying}
\newacronym{at}{AT}{ATtention}
\newacronym{auv}{AUV}{autonomous underwater vehicle}
\newacronym{awg}{AWG}{arbitrary waveform generator}
\newacronym{awgn}{AWGN}{additive white Gaussian noise}
\newacronym{baw}{BAW}{bulk acoustic wave}
\newacronym{bb}{BB}{base-band}
\newacronym{bc}{BC}{backscatter communication}
\newacronym{bcjr}{BCJR}{Bahl-Cocke-Jelinek-Raviv}
\newacronym{bd}{BD}{backscatter device}
\newacronym{be}{BE}{Belgium}
\newacronym{ber}{BER}{bit error rate}
\newacronym{bf}{BF}{beamforming}
\newacronym{bga}{BGA}{ball grid array}
\newacronym{bldc}{BLDC}{brushless DC}
\newacronym{bler}{BLER}{block error rate}
\newacronym{bod}{BOD}{Brown-Out Detection}
\newacronym{bom}{BOM}{bill of materials}
\newacronym{bpsk}{BPSK}{binary phase-shift keying}
\newacronym{brp}{BRP}{beam refinement process}
\newacronym{bs}{BS}{base station}
\newacronym{bu}{BU}{booster unit}
\newacronym{bw}{BW}{bandwidth}
\newacronym{ca}{CA}{carrier emitter}
\newacronym{cad}{CAD}{channel activity detection}
\newacronym{cars}{CARS}{calibration reference signal}
\newacronym{cbm}{CBM}{condition based maintenance}
\newacronym{cc}{CC}{constant current}
\newacronym{ccdf}{CCDF}{complementary cumulative distribution function}
\newacronym{ccnn}{CCNN}{circular convolutional neural network}
\newacronym{ccs}{CCS}{correlative channel sounder}
\newacronym{cdf}{CDF}{cumulative distribution function}
\newacronym{cdma}{CDMA}{code division-multiple access}
\newacronym{cdrx}{CDRX}{connected mode DRX}
\newacronym{ce}{CE}{coverage enhancement}
\newacronym{ced}{CED}{cumulative energy density}
\newacronym{ceofdm}{CE-OFDM}{constant-envelope OFDM}
\newacronym{cf}{CF}{cell-free}
\newacronym{cfmmimo}{CF-mMIMO}{cell-free massive MIMO}
\newacronym{cfo}{CFO}{carrier frequency offset}
\newacronym{cfr}{CFR}{channel frequency response}
\newacronym{cir}{CIR}{channel impulse response}
\newacronym{cla}{CLA}{closed-loop approach}
\newacronym{clk}{CLK}{clock}
\newacronym{cmos}{CMOS}{complementary metal oxide semiconductor}
\newacronym{cnn}{CNN}{convolutional neural network}
\newacronym{co}{CO}{combinatorial optimization}
\newacronym{cordic}{CORDIC}{coordinate rotation digital computer}
\newacronym{cost}{COST}{commercial off-the-shelf}
\newacronym{cots}{COTS}{commercial off-the-shelf}
\newacronym{cp}{CP}{cyclic prefix}
\newacronym{cpe}{CPE}{common phase error}
\newacronym{cpfsk}{CPFSK}{continuous phase frequency shift keying}
\newacronym{cpm}{CPM}{continuous phase modulation}
\newacronym{cpt}{CPT}{capacitive power transfer}
\newacronym{cpu}{CPU}{central-processing unit}
\newacronym{cpw}{CPW}{coplanar waveguide}
\newacronym{cqi}{CQI}{channel quality indicator}
\newacronym{cr}{CR}{coding rate}
\newacronym{crc}{CRC}{cyclic redundancy check}
\newacronym{crlb}{CRLB}{Cram\'er-Rao lower bound}
\newacronym{crs}{CRS}{cell reference signal}
\newacronym{cs}{CS}{compressed sensing}
\newacronym{cse}{CSE}{channel state estimation}
\newacronym{csi}{CSI}{channel state information}
\newacronym{csp}{CSP}{contact service point}
\newacronym{css}{CSS}{chirp spread spectrum}
\newacronym{cu}{CU}{central unit}
\newacronym{cv}{CV}{constant voltage}
\newacronym{cw}{CW}{continuous wave}
\newacronym{d2d}{D2D}{device-to-device}
\newacronym{dab}{DAB}{Digital Audio Broadcasting}
\newacronym{dac}{DAC}{digital-to-analog converter}
\newacronym{daq}{DAQ}{data acquisition system}
\newacronym{das}{DAS}{distributed antenna systems}
\newacronym{dbpsk}{DBPSK}{Differential Binary Phase Shift Keying}
\newacronym{dc}{DC}{direct current}
\newacronym{dcc}{DCC}{dynamic cooperation clustering}
\newacronym{ddc}{DDC}{digital down conversion}
\newacronym{de}{DE}{drain efficiency}
\newacronym{dft}{DFT}{discrete Fourier transform}
\newacronym{dftsofdm}{DFT-s-OFDM}{discrete Fourier Transform spread OFDM}
\newacronym{dftsofdmfdss}{DFT-s-OFDM-FDSS}{DFT-s-OFDM with frequency-domain spectral shaping}
\newacronym{dftsofdmfdssse}{DFT-s-OFDM-FDSS-SE}{DFT-s-OFDM with FDSS and spectral extension}
\newacronym{dl}{DL}{downlink}
\newacronym{dlc}{DLC}{Distributed Laser Charging}
\newacronym{dli}{DLI}{direct link interference}
\newacronym{dlt}{DLT}{Distributed Ledger Technology}
\newacronym{dm}{DM}{diffuse multipath}
\newacronym{dma}{DMA}{Direct Memory Access}
\newacronym{dmac}{DMAC}{Direct Memory Access Controller}
\newacronym{dmc}{DMC}{diffuse multipath component}
\newacronym{dmimo}{D-MIMO}{distributed MIMO}
\newacronym{dnn}{DNN}{deep neural network}
\newacronym{doa}{DOA}{direction-of-arrival}
\newacronym{dp}{DP}{differencial privacy}
\newacronym{dpd}{DPD}{digital pre-distortion}
\newacronym{dpdk}{DPDK}{Data Plane Development Kit}
\newacronym{dram}{DRAM}{dynamic random-access memory}
\newacronym{drcs}{$\Delta$RCS}{differential-radar cross section}
\newacronym{drx}{DRX}{Discontinuous Reception Mode}
\newacronym{dsb}{DSB}{double-sideband}
\newacronym{dsl}{DSL}{digital subscriber line}
\newacronym{dsp}{DSP}{digital signal processing}
\newacronym{dss}{DSS}{Dataset Storage Standard}
\newacronym{duc}{DUC}{digital up-converter}
\newacronym{dvb}{DVB}{Digital Video Broadcasting}
\newacronym{e2e}{E2E}{end-to-end}
\newacronym{easa}{EASA}{European Union Aviation Safety Agency}
\newacronym{ebg}{EBG}{electromagnetic bandgap}
\newacronym{ebw}{EBW}{excess bandwidth}
\newacronym{ec}{EC}{European Commission}
\newacronym{ecc}{ECC}{elliptic curve cryptography}
\newacronym{ecdf}{eCDF}{empirical cumulative distribution function}
\newacronym{ecdlp}{ECDLP}{elliptic curve discrete logarithm problem}
\newacronym{ecsp}{ECSP}{edge computing service point}
\newacronym{edlc}{EDLC}{electrostatic double-layer capacitors}
\newacronym{edrx}{eDRX}{Extended Discontinuous Reception Mode}
\newacronym{ee}{EE}{energy efficiency}
\newacronym{egprs}{EGPRS}{Enhanced Data Rates for GSM Evolution}
\newacronym{eh}{EH}{energy harvesting}
\newacronym{eirp}{EIRP}{equivalent isotropically radiated power}
\newacronym{em}{EM}{electromagnetic}
\newacronym{embb}{eMBB}{enhanced Mobile Broadband}
\newacronym{en}{EN}{energy neutral}
\newacronym{end}{END}{energy-neutral device}
\newacronym{enob}{ENOB}{effective number of bits}
\newacronym{eol}{EoL}{end of life}
\newacronym{ep}{EP}{energy profiler}
\newacronym{epd}{EPD}{electronic paper display}
\newacronym{epu}{EPU}{edge processing unit}
\newacronym{er}{ER}{Energy Receiver}
\newacronym{erp}{ERP}{effective radiation power}
\newacronym[plural=ESCs,firstplural=electronic speed controllers (ESCs)]{esc}{ESC}{electronic speed control}
\newacronym{esd}{ESD}{electrostatic discharge}
\newacronym{esl}{ESL}{electronic shelf label}
\newacronym{esr}{ESR}{equivalent series resistance}
\newacronym{et}{ET}{Energy Transmitter}
\newacronym{etsi}{ETSI}{European Telecommunications Standards Institute}
\newacronym{evd}{EVD}{eigenvalue decomposition}
\newacronym{evm}{EVM}{Error Vector Magnitude}
\newacronym{ewlb}{eWLB}{Embedded Wafer Level Ball Grid Array}
\newacronym{fa}{FA}{federation anchor}
\newacronym{fair}{FAIR}{Findability, Accessibility, Interoperability, and Reuse of digital assets}
\newacronym{fcf}{FCF}{frequency correlation function}
\newacronym{fd}{FD}{front-haul distance}
\newacronym{fdd}{FDD}{frequency-division duplexing}
\newacronym{fde}{FDE}{frequency domain equalizer}
\newacronym{fdm}{FDM}{frequency-division multiplexing}
\newacronym{fdma}{FDMA}{frequency division multiple access}
\newacronym{fdss}{FDSS}{frequency-domain spectral shaping}
\newacronym{fem}{FEM}{finite element analysis}
\newacronym{fembb}{feMBB}{further enhanced mobile broadband}
\newacronym{fft}{FFT}{fast Fourier transform}
\newacronym{fh}{FH}{fronthaul}
\newacronym{fhss}{FHSS}{frequency hopping spread spectrum}
\newacronym{fifo}{FIFO}{First In, First Out}
\newacronym{fim}{FIM}{Fisher information matrix}
\newacronym{fits}{FITS}{Flexible Image Transport System}
\newacronym{fl}{FL}{federated learning}
\newacronym{fom}{FoM}{figure of merit}
\newacronym{fov}{FOV}{field of view}
\newacronym{fpga}{FPGA}{field-programmable gate array}
\newacronym{fr1}{FR1}{frequency range 1}
\newacronym{fr2}{FR2}{frequency range 2}
\newacronym{fram}{FRAM}{Ferroelectric Random Access Memory}
\newacronym{fsk}{FSK}{frequency shift keying}
\newacronym{fspl}{FSPL}{free space path loss}
\newacronym{fss}{FSS}{frequency selective surface}
\newacronym{gb}{GB}{grant-based}
\newacronym{gdpr}{GDPR}{general data protection regulation}
\newacronym{gf}{GF}{grant-free}
\newacronym{gmsk}{GMSK}{Gaussian minimum-shift keying}
\newacronym{gnb}{gNB}{Next Generation Node B}
\newacronym{gni}{GNI}{gross national income}
\newacronym{gnn}{GNN}{graph neural network}
\newacronym{gnss}{GNSS}{global navigation satellite system}
\newacronym{gpclk}{GPCLK}{general purpose clock}
\newacronym{gpio}{GPIO}{General-Purpose Input/Output}
\newacronym{gpl}{GPL}{GNU General Public License}
\newacronym{gprs}{GPRS}{General Packet Radio Services}
\newacronym{gps}{GPS}{Global Positioning System}
\newacronym{gpu}{GPU}{graphical processing unit}
\newacronym{grc}{GRC}{GNU Radio Companion}
\newacronym{gscm}{GSCM}{geometry‐based stochastic model}
\newacronym{gsm}{GSM}{Global System for Mobile Communications}  %
\newacronym{gspm}{GSpM}{Generalized spatial modulation}
\newacronym{gwp}{GWP}{Global Warming Potential}
\newacronym{harq}{HARQ}{hybrid automatic repeat request}
\newacronym{hat}{HAT}{hardware attached on top}
\newacronym{hcs}{HCS}{human-centric services}
\newacronym{hdf5}{HDF5}{Hierarchical Data Format version 5}
\newacronym{hfss}{HFSS}{High Frequency Simulator Software}
\newacronym{hmd}{HMD}{head-mounted display}
\newacronym{hpbm}{HPBM}{half power beam width}
\newacronym{HyMPRo}{HyMPRo}{Hybrid Multi-Path Routing algorithm}
\newacronym{i.i.d.}{i.i.d.}{independent and identically distributed}
\newacronym{i2c}{I2C}{Inter-Integrated Circuit}
\newacronym{iaq}{IAQ}{Indoor Air Quality}
\newacronym{ib}{IB}{in-band}
\newacronym{ibbc}{IBBC}{inter-band beam configuration}
\newacronym{ibo}{IBO}{input back-off}
\newacronym{ic}{IC}{integrated circuit}
\newacronym{iccs}{ICCS}{Ilmsens correlative channel sounder}
\newacronym{ici}{ICI}{intercarrier interference}
\newacronym{icnirp}{ICNIRP}{International Commission on Non-Ionizing Radiation Protection}
\newacronym{id}{ID}{information decoding}
\newacronym{idft}{IDFT}{inverse discrete Fourier transform}
\newacronym{idxm}{IDXM}{index modulation}
\newacronym{if}{IF}{intermediate-frequency}
\newacronym{ifft}{IFFT}{inverse fast-Fourier-transform}
\newacronym{iid}{i.i.d.}{independently and identically distributed}
\newacronym{iis}{IIS}{integrated information system}
\newacronym{im}{IM}{intermodulation}
\newacronym{imd}{IMD}{intermodulation distortion}
\newacronym{imu}{IMU}{inertial measurement unit}
\newacronym{inh}{InH}{indoor hotspot office}
\newacronym{io}{IO}{input/output}
\newacronym{ioe}{IoE}{Internet of Everything}
\newacronym{iot}{IoT}{Internet of Things}
\newacronym{ipt}{IPT}{inductive power transfer}
\newacronym{ipy}{IPY}{Interventions per Year}
\newacronym{iq}{IQ}{in-phase and quadrature}
\newacronym{iqi}{IQI}{IQ imbalance}
\newacronym{ir}{IR}{infrared}
\newacronym{isi}{ISI}{intersymbol interference}
\newacronym{ism}{ISM}{industrial, scientific and medical}
\newacronym{isp}{ISP}{internet service provider}
\newacronym{itu}{ITU}{International Telecommunication Union}
\newacronym{jesd}{JESD}{Joint Electron Devices Engineering Council}
\newacronym{jfet}{JFET}{junction field effect transistor}
\newacronym{kpi}{KPI}{key performance indicator}
\newacronym{ktofdm}{KT-DFT-s-OFDM}{known-tail-DFT-s-OFDM}
\newacronym{kvi}{KVI}{key value indicator}
\newacronym{larva}{LARVA}{LARge Virtual Array}
\newacronym{lca}{LCA}{life cycle assessment}
\newacronym{lco}{LCO}{lithium cobalt oxide}
\newacronym{ldo}{LDO}{Low-dropout voltage regulator}
\newacronym{ldpc}{LDPC}{low-density parity-check}
\newacronym{led}{LED}{Light Emitting Diode}
\newacronym{less}{LESS}{Low Energy Scheduler Solution}
\newacronym{lfp}{LFP}{lithium iron phosphate}
\newacronym{lib}{LIB}{Lithium-Ion Battery}
\newacronym{lic}{LIC}{lithium-ion capacitor}
\newacronym{lid}{LID}{Lithium Iron Disulfide}
\newacronym{lidar}{LiDAR}{light detection and ranging}
\newacronym{liion}{Li-ion}{lithium-ion}
\newacronym{lipo}{LiPo}{lithium polymer}
\newacronym{lis}{LIS}{large intelligent surface}
\newacronym{llh}{LLH}{log-likelihood}
\newacronym{lls}{LLS}{link-level simulation}
\newacronym{lmd}{LMD}{Lithium Manganese Dioxide}
\newacronym{lmmse}{LMMSE}{least minimum mean square error}
\newacronym{lmo}{LMO}{lithium ion manganese oxide}
\newacronym{lna}{LNA}{low-noise amplifier}
\newacronym{lo}{LO}{local oscillator}
\newacronym{lora}{LoRa}{long range}
\newacronym{lorawan}{LoRaWAN}{long-range wide-area network}
\newacronym{los}{LoS}{line-of-sight}
\newacronym{lp}{LP}{linear programming}
\newacronym{lpf}{LPF}{low-pass filter}
\newacronym{lpt}{LPT}{laser power transfer}
\newacronym{lpwa}{LPWA}{Low Power Wide Area}
\newacronym{lpwan}{LPWAN}{low-power wide-area network}
\newacronym{lpwans}{LPWANs}{Low-Power Wide-Area Networks}
\newacronym{lqi}{LQI}{link quality indicator}
\newacronym{lrelu}{LReLU}{leaky rectified linear unit}
\newacronym{lrt}{LRT}{likelihood-ratio test}
\newacronym{ls}{LS}{least squares}
\newacronym{lsa}{LSA}{large synthetic array}
\newacronym{lsf}{LSF}{large-scale fading}
\newacronym{lsfc}{LSFC}{large-scale fading component}
\newacronym{lstm}{LSTM}{Long Short-Term Memory}
\newacronym{ltc}{LTC}{lithium thionyl chloride}
\newacronym{lte}{LTE}{Long Term Evolution}
\newacronym{lti}{LTI}{linear time-invariant}
\newacronym{lto}{LTO}{lithium titanate}
\newacronym{lusta}{LUSTA 5G }{Logistique mUltimodale Sécuritaire Téléopérée \& Autonome 5G}
\newacronym{m2m}{M2M}{machine to machine}
\newacronym{mac}{MAC}{Medium Access Control}
\newacronym{mate}{MATE}{millimeter-wave MIMO testbed}
\newacronym{mc}{MC}{Monte Carlo}
\newacronym{mcl}{MCL}{Maximum Coupling Loss}
\newacronym{mcs}{MCS}{modulation and coding scheme}
\newacronym{mcu}{MCU}{microcontroller unit}
\newacronym{mec}{MEC}{multi-access edge computing}
\newacronym{mems}{MEMS}{micro-electromechanical systems}
\newacronym{mf}{MF}{matched filter}
\newacronym{mimo}{MIMO}{multiple-input multiple-output}
\newacronym{miso}{MISO}{multiple-input single-output}
\newacronym{ml}{ML}{machine learning}
\newacronym{mlp}{MLP}{multilayer perceptron}
\newacronym{mmic}{MMIC}{monolithic microwave integrated circuit}
\newacronym{mmimo}{mMIMO}{massive MIMO}
\newacronym{mmse}{MMSE}{minimum mean square error}
\newacronym{mmtc}{mMTC}{massive machine-typed communication}
\newacronym{mmwave}{mmWave}{millimeter wave}
\newacronym{mn}{MN}{matching network}
\newacronym{mosfet}{MOSFET}{metal-oxide semiconductor field effect transistor}
\newacronym{mpc}{MPC}{multipath component}
\newacronym{mppt}{MPPT}{maximum power point tracking}
\newacronym{mr}{MR}{maximum ratio}
\newacronym{mrc}{MRC}{maximum ratio combining}
\newacronym{mrc_em}{MRC}{maximum ratio combining}
\newacronym{mrc_EM}{MRC}{Magnetic Resonance Coupling}
\newacronym{mrt}{MRT}{maximum ratio transmission}
\newacronym{mse}{MSE}{mean square error}
\newacronym{msk}{MSK}{Minimum-Shift Keying}
\newacronym{mtc}{MTC}{Machine-Type Communication}
\newacronym{multi-rat}{Multi-RAT}{multiple radio access technology}
\newacronym{multirat}{Multi-RAT}{Multiple Radio Access Technology}
\newacronym{music}{MUSIC}{MUltiple SIgnal Classification}
\newacronym{navauwall}{NAVAUWALL}{AUtomated NAVigation in WALLonia}
\newacronym{nb}{NB}{narrowband}
\newacronym{nbiot}{NB-IoT}{narrowband IoT}
\newacronym{nca}{NCA}{nickel cobalt aluminum}
\newacronym{netcdf}{NetCDF}{Network Common Data Form}
\newacronym{nf}{NF}{noise figure}
\newacronym{nfc}{NFC}{near-field communication}
\newacronym{nfv}{NFV}{network function virtualization}
\newacronym{ngmn}{NGMN}{Next Generation Mobile Networks }
\newacronym{ni}{NI}{National Instruments}
\newacronym{nicd}{NiCd}{nikkel cadmium}
\newacronym{nimh}{NiMH}{nikkel metal hydride}
\newacronym{nlos}{NLoS}{non-line-of-sight}
\newacronym{nmc}{NMC}{nickel manganese cobalt}
\newacronym{nmos}{nMOS}{n-channel metal-oxide semiconductor}
\newacronym{nn}{NN}{neural network}
\newacronym{nnls}{NNLS}{non-negative least squares}
\newacronym{noma}{NOMA}{non-orthogonal multiple access}
\newacronym{np}{NP}{Neyman-Pearson}
\newacronym{npbch}{NPBCH}{Narrowband Physical Broadcast Channel}
\newacronym{npss}{NPSS}{Narrow Band Primay Synchronization Signal}
\newacronym{nr}{NR}{New Radio}
\newacronym{nrs}{NRS}{Narrow Band Reference Signal}
\newacronym{nsss}{NSSS}{Narrowband Secondary Synchronization Signal}
\newacronym{ntp}{NTP}{network time protocol}
\newacronym{oai}{OAI}{OpenAirInterface} %
\newacronym{obw}{OBW}{occupied bandwidth}
\newacronym{ofdm}{OFDM}{orthogonal frequency-division multiplexing}
\newacronym{ofdma}{OFDMA}{orthogonal frequency-division multiple access}
\newacronym{ofdmim}{OFDM-IM}{OFDM with index modulation}
\newacronym{olos}{OLoS}{obstructed-line-of-sight}
\newacronym{oma}{OMA}{orthogonal multiple access}
\newacronym{oob}{OOB}{out-of-band}
\newacronym{ook}{OOK}{on-off keying}
\newacronym{opbo}{OPBO}{output power backoff}
\newacronym{oran}{O-RAN}{open radio-access network}
\newacronym{os}{OS}{operating system}
\newacronym{ota}{OTA}{over-the-air}
\newacronym{otaa}{OTAA}{over-the-air authentication}
\newacronym{p1}{P1}{Phase 1}
\newacronym{p2}{P2}{Phase 2}
\newacronym{p2p}{P2P}{point-to-point}
\newacronym{pa}{PA}{power amplifier}
\newacronym{pae}{PAE}{power-added efficiency}
\newacronym{pam}{PAM}{pulse amplitude modulation}
\newacronym{pana}{PanA}{Panel A}
\newacronym{panb}{PanB}{Panel B}
\newacronym{papr}{PAPR}{peak-to-average power ratio}
\newacronym{pc}{PC}{pilot count}
\newacronym{pcb}{PCB}{printed circuit board}
\newacronym{pcg}{PCG}{power consumption gain}
\newacronym{pcie}{PCIe}{Peripheral Component Interconnect Express}
\newacronym{pcsi}{PCSI}{perfect channel state information}
\newacronym{pd}{PD}{powered device}
\newacronym{pdcch}{PDCCH}{physical downlink control channel}
\newacronym{pdf}{PDF}{probability density function}
\newacronym{pdp}{PDP}{power delay profile}
\newacronym{pdsch}{PDSCH}{physical downlink shared channel}
\newacronym{pe}{PE}{processing element}
\newacronym{peb}{PEB}{positioning error bound}
\newacronym{per}{PER}{packet error rate}
\newacronym{pet}{PET}{privacy enhancing technology}
\newacronym{pg}{PG}{path gain}
\newacronym{pgd}{PGD}{proximal gradient descent}
\newacronym{phy}{PHY}{physical}
\newacronym{pki}{PKI}{public key infrastucture}
\newacronym{pl}{PL}{path loss}
\newacronym{pla}{PLA}{physically large array}
\newacronym{pll}{PLL}{phase-locked loop}
\newacronym[plural=PMs,firstplural=person months (PMs)]{pm}{PM}{person month}
\newacronym{pmf}{PMF}{polymer microwave fiber}
\newacronym{pmu}{PMU}{power management unit}
\newacronym{pn}{PN}{pseudo-noise}
\newacronym{po}{PO}{phase offset}
\newacronym{poc}{PoC}{proof of concept}
\newacronym{poe}{PoE}{power-over-Ethernet}
\newacronym{pps}{1PPS}{pulse per second}
\newacronym{pr}{PR}{phase reversal}
\newacronym{prb}{PRB}{Physical Resource Block}
\newacronym{prbs}{PRBs}{Physical Resource Blocks}
\newacronym{prs}{PRS}{Peripheral Reflex System}
\newacronym{ps}{PS}{Processing System}
\newacronym{psd}{PSD}{power spectral density}
\newacronym{pse}{PSE}{power sourcing equipment}
\newacronym{psk}{PSK}{phase shift keying}
\newacronym{psm}{PSM}{power saving mode}
\newacronym{pss}{PSS}{primary synchronisation signal}
\newacronym{ptp}{PTP}{precision-time protocol}
\newacronym{ptrs}{PTRS}{Phase-Tracking Reference Signals}
\newacronym{ptw}{PTW}{paging time window}
\newacronym{pv}{PV}{photovoltaic}
\newacronym{pw}{PW}{planar wavefront}
\newacronym{pwm}{PWM}{pulse width modulation}
\newacronym{qam}{QAM}{quadrature amplitude modulation}
\newacronym{qos}{QoS}{quality-of-service}
\newacronym{qpsk}{QPSK}{quadrature phase-shift keying}
\newacronym{qrrls}{QR-RLS}{QR decomposition based recursive least squares}
\newacronym{quadriga}{QuaDRiGa}{QUAsi Deterministic RadIo channel GenerAtor}
\newacronym{ra}{RA}{Random Access}
\newacronym{ram}{RAM}{random-access memory}
\newacronym{ran}{RAN}{radio access network}
\newacronym{rar}{RAR}{Random Access Response}
\newacronym{rat}{RAT}{radio access technology}
\newacronym{rb}{Rb}{Rubidium}
\newacronym{rbs}{RBS}{radio base station}
\newacronym{rbw}{RBW}{resolution bandwidth}
\newacronym{rc}{RC}{raised-cosine}
\newacronym{rcs}{RCS}{radar cross section}
\newacronym{rdl}{RDL}{redistribution layer}
\newacronym{re}{RE}{radio element}
\newacronym{relu}{ReLU}{rectified linear unit}
\newacronym{rf}{RF}{radio frequency}
\newacronym{rfeh}{RFEH}{radio frequency energy harvesting}
\newacronym{rfic}{RFIC}{radio-frequency integrated circuit}
\newacronym{rfid}{RFID}{radio frequency identification}
\newacronym{rfpt}{RFPT}{radio frequency power transfer}
\newacronym{rfsoc}{RFSoC}{Radio Frequency System-on-Chip}
\newacronym{rir}{RIR}{room impulse response}
\newacronym{ris}{RIS}{reflective intelligent surface}%
\newacronym{rllmtc}{RLLMTC}{reliable low latency machine type communication}
\newacronym{rls}{RLS}{recursive least squares}
\newacronym{rms}{RMS}{root-mean-square}
\newacronym{rmse}{RMSE}{root-mean-square error}
\newacronym{rmt}{RMT}{random matrix theory}
\newacronym{rof}{RoF}{radio-over-fiber}
\newacronym{ros}{ROS}{robot operating system}
\newacronym{rpi}{RPi}{Raspberry Pi}
\newacronym{rrc}{RRC}{Radio Resource Connection}
\newacronym{rreq}{RREQ}{route request packet}
\newacronym{rsrp}{RSRP}{Reference Signals Received Power}
\newacronym{rsrq}{RSRQ}{Reference Signal Received Quality}
\newacronym{rss}{RSS}{received signal strength}
\newacronym{rssi}{RSSI}{received signal strength indicator}
\newacronym{rtc}{RTC}{real time clock}
\newacronym{rtf}{RTF}{reader talks first}
\newacronym{rtk}{RTK}{real time kinematics}
\newacronym{rts}{RTS}{ray tracing simulator}
\newacronym{ru}{RU}{Radio Unit}
\newacronym{rv}{RV}{random variable}
\newacronym{rw}{RW}{RadioWeaves}
\newacronym{rx}{RX}{receiver}
\newacronym{rzf}{RZF}{regularized zero forcing}
\newacronym{rt}{RT}{ray-tracer}
\newacronym{s-parameter}{S-parameter}{scattering parameter}
\newacronym{sa}{SA}{synchronization anchor}
\newacronym{sa5}{SA}{Stand Alone}
\newacronym{sar}{SAR}{specific absorption rate}
\newacronym{sbl}{SBL}{sparse Bayesian learning}
\newacronym{sc}{SC}{single carrier}
\newacronym{scfde}{SC-FDE}{single-carrier modulation with frequency-domain-equalization}
\newacronym{scfdma}{SCFDMA}{single-carrier frequency division multiple access}
\newacronym{scs}{SCS}{sub-carrier spacing}
\newacronym{sdg}{SDG}{Sustainable Development Goal}
\newacronym{sdm}{SDM}{sigma-delta modulator}
\newacronym{sdma}{SDMA}{spatial-division multiple access}
\newacronym{sdn}{SDN}{software-defined network}
\newacronym{sdof}{SDoF}{sigma-delta over fiber}
\newacronym{sdr}{SDR}{software-defined radio}
\newacronym{se}{SE}{spectral efficiency}
\newacronym{sei}{SEI}{specific emitter identification}
\newacronym{ser}{SER}{symbol-error rate}
\newacronym{sf}{SF}{spreading factor}
\newacronym{sfn}{SFN}{single frequency network}
\newacronym{sfp}{SFP}{small form-factor pluggable}
\newacronym{sha}{SHA}{Secure Hash Algorithm}
\newacronym{SigMF}{SigMF}{Signal Metadata Format}
\newacronym{simo}{SIMO}{single-input multiple-output}
\newacronym{sinr}{SINR}{signal-to-interference-plus-noise ratio}
\newacronym{siso}{SISO}{single-input single-output}
\newacronym{slam}{SLAM}{simultaneous localization and mapping}
\newacronym{slc}{SLC}{spatial leakage suppression}
\newacronym{slerp}{SLERP}{spherical linear interpolation}
\newacronym{sma}{SMA}{SubMiniature version A}
\newacronym{smc}{SMC}{specular multipath component}
\newacronym{smps}{SMPS}{switched mode power supply}
\newacronym{sndr}{SNDR}{signal-to-noise-and-distortion ratio}
\newacronym{snidr}{SNIDR}{signal-to-noise-and-interference-and-distortion ratio}
\newacronym{snir}{SNIR}{signal-to-interference-plus-noise ratio}
\newacronym{snr}{SNR}{signal-to-noise ratio}
\newacronym{soc}{SoC}{state of charge}
\newacronym{SoC}{SOC}{System on Chip}
\newacronym{sp}{SP}{service point}
\newacronym{spdt}{SPDT}{single pole double throw}
\newacronym{spi}{SPI}{Serial Peripheral Interface}
\newacronym{spst}{SPST}{single pole single throw}
\newacronym{sram}{SRAM}{static random-access memory}
\newacronym{srd}{SRD}{short-range device}
\newacronym{srls}{SRLS}{standard recursive least squares}
\newacronym{srs}{SRS}{Sounding Reference Signal}
\newacronym{ssb}{SSB}{synchronisation signal block}
\newacronym{ssd}{SSD}{solid state drive}
\newacronym{ssq}{SSQ}{simulator sickness questionnaire}
\newacronym{steam}{STEAM}{science, technology, engineering, the arts, and mathematics}
\newacronym{svd}{SVD}{singular value decomposition}
\newacronym{sw}{SW}{spherical wavefront}
\newacronym{swipt}{SWIPT}{simultaneous wireless information and power transfer}
\newacronym{synce}{SyncE}{Synchronous Ethernet}
\newacronym{ta}{TA}{timing advance}
\newacronym{tau}{TAU}{tracking area update}
\newacronym{tcer}{TCER}{transported to consumed energy ratio}
\newacronym{tcp}{TCP}{Transmission Control Protocol}
\newacronym{tcxo}{TCXO}{temperature compensated crystal oscillator}
\newacronym{tdce}{TD-CE}{time-domain compression and expansion}
\newacronym{tdd}{TDD}{time division duplexing}
\newacronym{tdma}{TDMA}{time division-multiple access}
\newacronym{tdoa}{TDOA}{time-difference-of-arrival}
\newacronym{thz}{THz}{Terahertz}
\newacronym{to}{TO}{timing offset}
\newacronym{toa}{TOA}{time-of-arrival}
\newacronym{tof}{ToF}{time-of-flight}
\newacronym{tosm}{TOSM}{through-open-short-match}
\newacronym{tpms}{TPMS}{Tire-Pressure Monitoring System}
\newacronym{trl}{TRL}{technology readyness level}
\newacronym{trp}{TRP}{Transmission Reception Point}
\newacronym{tsn}{TSN}{time-sensitive networking}
\newacronym{ttf}{TTF}{tag talks first}
\newacronym{ttff}{TTFF}{Time To First Fix}
\newacronym{tti}{TTI}{transmission time interval}
\newacronym{ttm}{TTM}{time to market}
\newacronym{ttn}{TTN}{The Things Network}
\newacronym{tx}{TX}{transmitter}
\newacronym{tdl}{TDL}{tapped delay line}
\newacronym{uart}{UART}{Universal Asynchronous Receiver/Transmitter}
\newacronym{uav}{UAV}{unmanned aerial vehicle}
\newacronym{uc}{UC}{use case}
\newacronym{ucie}{UCIe}{Universal Chiplet Interconnect Express}
\newacronym{udp}{UDP}{User Datagram Protocol}
\newacronym{ue}{UE}{user equipment}
\newacronym{ugv}{UGV}{unmanned ground vehicle}
\newacronym{uhd}{UHD}{USRP hardware driver}
\newacronym{uhf}{UHF}{ultra-high frequency}
\newacronym{ul}{UL}{uplink}
\newacronym{ula}{ULA}{uniform linear array}
\newacronym{uMIMO}{$\mu$-MIMO}{ultra-massive Multiple-Input Multiple-Output}
\newacronym{ummtc}{umMTC}{ultra massive machine type communication}
\newacronym{un}{UN}{United Nations}
\newacronym{unow}{uNOW}{unified non-orthogonal waveform}
\newacronym{upa}{UPA}{uniform planar array}
\newacronym{ura}{URA}{uniform rectangular array}
\newacronym{urllc}{URLLC}{ultra-reliable low-latency communications}
\newacronym{usrp}{USRP}{universal software radio peripheral}
\newacronym{uv}{UV}{unmanned vehicle}
\newacronym{uw}{UW}{unique word}
\newacronym{uwb}{UWB}{ultrawideband}
\newacronym{uwofdm}{UW-DFT-s-OFDM}{unique word DFT-s-OFDM}
\newacronym{v2v}{V2V}{vehicle-to-vehicle}
\newacronym{vco}{VCO}{voltage-controlled oscillator}
\newacronym{vep}{VEP}{virtual edge platform}
\newacronym{vlc}{VLC}{visible light communication}
\newacronym{vlp}{VLP}{visible light positioning}
\newacronym{vna}{VNA}{vector network analyzer}
\newacronym{voc}{VOC}{Voltatile Organic Compound}
\newacronym{votable}{VOTable}{Virtual Observatory Table}
\newacronym{vr}{VR}{virtual reality}
\newacronym{vuca}{VUCA}{volatile, uncertain, complex and ambiguous}
\newacronym{wb}{WB}{wideband}
\newacronym{wban}{WBAN}{wireless body area network}
\newacronym{wd}{WD}{Wireless distance}
\newacronym{wimax}{WiMAX}{Worldwide Interoperability for Microwave Access}
\newacronym{wlan}{WLAN}{wireless LAN}
\newacronym[plural=WPs,firstplural=work packages (WPs)]{wp}{WP}{work package}
\newacronym{wpt}{WPT}{wireless power transfer}
\newacronym{wr}{WR}{White Rabbit}
\newacronym{wrsn}{WRSN}{wireless rechargeable sensor network}
\newacronym{wsn}{WSN}{wireless sensor network}
\newacronym{xets}{XETS}{cross exponentially tapered slot}
\newacronym{xlmimo}{XL-MIMO}{extremely large-scale MIMO}
\newacronym{xr}{XR}{extended reality}
\newacronym{z3ro}{Z3RO}{zero third-order distortion}
\newacronym{zf}{ZF}{zero-forcing}
\newacronym{zmcscg}{ZMCSCG}{zero mean circularly symmetric complex Gaussian}
\newacronym{zmq}{ZMQ}{ZeroMQ}
\newacronym{ztofdm}{ZT-DFT-s-OFDM}{zero-tail DFT-s-OFDM}
\newacronym{llr}{LLR}{log-likelihood ratio}
\newacronym{nmse}{NMSE}{normalised mean square error}
  \glsdisablehyper                %

\input{auto_names}                  %

\usepackage{orcidlink}
\AtBeginBibliography{\footnotesize}

\def\BibTeX{{\rm B\kern-.05em{\sc i\kern-.025em b}\kern-.08em
    T\kern-.1667em\lower.7ex\hbox{E}\kern-.125emX}}

\definecolor{color1}{HTML}{1b9e77}
\definecolor{color2}{HTML}{d95f02}
\definecolor{color3}{HTML}{7570b3}
\definecolor{color4}{HTML}{e7298a}
\definecolor{color5}{HTML}{66a61e}
\definecolor{color6}{HTML}{e6ab02}
\definecolor{color7}{HTML}{a6761d}
\definecolor{color8}{HTML}{666666}

\colorlet{AMBgreen}{color1}
\colorlet{AMBdarkgreen}{color1}
\colorlet{AMBlightgreen}{color1}
\definecolor{darkgray176}{RGB}{176,176,176}

\usepackage{tikz}
\usepackage{pgfplots}   %
\pgfplotsset{compat=newest}
\usetikzlibrary{plotmarks}
\usetikzlibrary{arrows.meta}
\usetikzlibrary{arrows}
\usetikzlibrary{calc,fadings,decorations.pathreplacing}
\usetikzlibrary{fit,backgrounds}  %

\usetikzlibrary{shapes,arrows,fit,positioning}
\usetikzlibrary{shapes}
\usetikzlibrary{spy}
\usetikzlibrary{circuits}
\usetikzlibrary{arrows}

\pgfplotsset{
    every axis/.append style={
        title style={draw=none},
        label style={font=\small},
        legend style={
            fill opacity=0.8,
            nodes={scale=0.8, transform shape}, {draw=none}
        },
        tick align=outside,
        tick pos=left,
        x grid style={darkgray176},
        xtick style={color=black},
        y grid style={darkgray176},
        ytick style={color=black},
        grid=both,
    },
    every axis plot/.append style={
        line width=2.0pt,
    },
}

\tikzset{%
  >=latex,
  inner sep=0pt,%
  outer sep=2pt,%
  mark coordinate/.style={inner sep=0pt,outer sep=0pt,minimum size=3pt,
  fill=black,circle}%
}

\usetikzlibrary{datavisualization.polar}

\usepackage[font=small]{caption}
\usepackage{subcaption}

\usepackage{fontawesome}
\usetikzlibrary {patterns,patterns.meta}
\usetikzlibrary{arrows.meta,positioning,calc}
\usetikzlibrary{intersections}
\usepackage{placeins}

\newcommand{\vect}[1]{\boldsymbol{\mathrm{#1}}}
\newcommand{\mat}[1]{\boldsymbol{\mathrm{#1}}}

\newcommand{\tr}[1]{\mathrm{tr}\left(#1\right)}
\newcommand{\diag}[1]{\mathrm{diag}\left(#1\right)}

\newcommand{\expt}[1]{\mathbb{E}\left(#1\right)}

\usetikzlibrary{external}

\hypersetup{
    pdfencoding=auto,
    psdextra,
    hidelinks,
    bookmarks=false,
    draft=true, 
}

\usepackage[
  paper=letterpaper,
  top=0.7in,
  bottom=1.0in,
  left=0.75in,
  right=0.75in,
  headheight=0pt,
  headsep=0pt,
  footskip=0.5in
]{geometry}

\DeclareUnicodeCharacter{0308}{\color{red} FIX ME!!!!}

\begin{document}

\title{An Open-Source Hardware-Aware Sub-THz Radio-Stripe Simulator
\thanks{This research was partially funded by 6GTandem, supported by the Smart Networks and Services Joint Undertaking (SNS JU) under the European Union's Horizon Europe research and innovation programme under Grant Agreement No 101096302.}%
\thanks{For the purpose of open access, the author has applied a CC BY public copyright license to any Author Accepted Manuscript version arising from this submission.} %
\thanks{All the relevant code for both the python simulator and the \gls{rt} module can be found on Github. \url{https://github.com/6GTandem}}
}

\author{
\IEEEauthorblockN{%
Tijl Schepens\,\orcidlink{0009-0007-7177-4141}\IEEEauthorrefmark{1},
Thomas Feys\,\orcidlink{0000-0002-5369-4670}\IEEEauthorrefmark{1},
Thomas Eriksson\,\orcidlink{0000-0002-2087-7227}\IEEEauthorrefmark{2},
Gilles Callebaut\,\orcidlink{0000-0003-2413-986X}\IEEEauthorrefmark{1},
}
\IEEEauthorblockA{\IEEEauthorrefmark{1}%
Department of Electrical Engineering, KU Leuven, Belgium}
\IEEEauthorblockA{\IEEEauthorrefmark{2}%
Department of Electrical Engineering, Chalmers University of Technology, Gothenburg, Sweden}}

\maketitle

\begin{abstract}
Sub-\acrlong{thz} radio-stripe and distributed MIMO architectures promise extreme spatial reuse and multi-GHz bandwidths, but the cascaded fiber front-haul and RF hardware impairments strongly shape end-to-end performance.
This paper presents an open-source, configuration-driven simulator that models the full waveform-level signal chain from CP-OFDM baseband generation in the \acrlong{cu}, through measurement-parameterized \acrlong{pmf} and coupler links, to booster/active \glspl{ru} with configurable nonlinearity, noise, \acrlong{iq} imbalance, and oscillator phase noise and \acrlong{cfo}.
Wireless propagation is supported via lightweight deterministic and stochastic per-subcarrier channel models as well as site-specific ray-tracing datasets generated with a companion Sionna \acrlong{rt} module.
The simulator exports intermediate waveforms and system metrics (e.g., \acrlong{nmse}, \acrlong{sndr}, \acrlong{ber}) to enable reproducible studies of impairment accumulation, calibration, and algorithmic choices such as \gls{ru} selection and beam management.
\end{abstract}


\begin{figure*}[ht!]
    \centering
    \includegraphics[width=\linewidth]{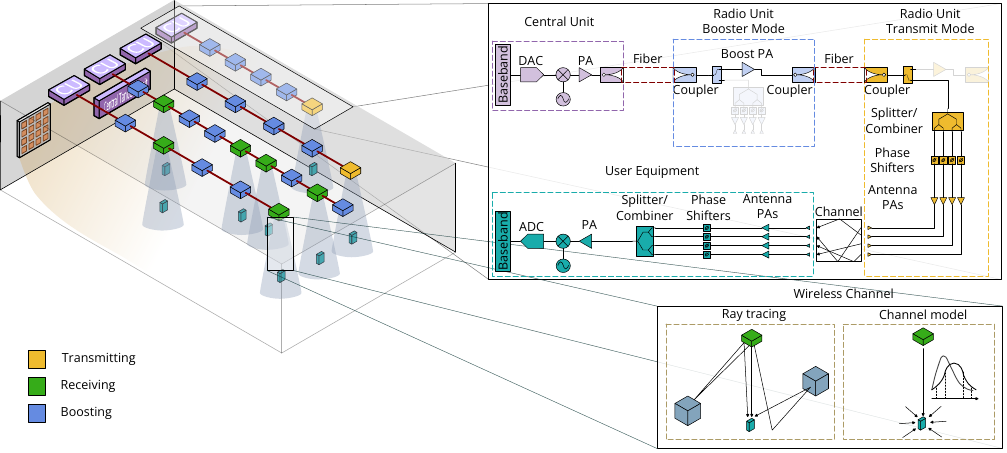}
    \caption{A  deployment of distributed low-complexity sub-Thz
      \glspl{ru} and sub-10GHz with dual-frequency operation. Block diagram of the
      sub-Thz fiber-based infrastructure (the stripe), consisting of a
      \gls{cu}, and multiple \glspl{ru} connected with \glspl{pmf}. A \gls{ru} can be configured to receive signals transmitted by a \gls{ue} (green), transmit over the air towards a \gls{ue} (orange) or amplify the signal (blue) to be carried further over the \gls{pmf} to account for the losses. 
      }
      \label{fig:Tandem_concepts}
\end{figure*}

\section{Introduction}

\glsresetall
6GTandem investigates dual-frequency distributed \gls{mimo} architectures combining sub-\SI{10}{\giga\hertz} and sub-\gls{thz} operation to enable high-capacity \gls{6g} services.
In the sub-\gls{thz} branch, the \emph{radio-stripe} concept distributes multiple \glspl{ru} along a fiber (\gls{pmf}/sub-\gls{thz} fiber), such that only a subset of \glspl{ru} actively transmits/receives while the remaining units boost and forward signals over the fiber.
This architecture creates tight coupling between system-level decisions (active \gls{ru} selection, beamforming, mobility) and hardware-level effects (fiber frequency selectivity, insertion losses, amplifier compression, oscillator phase noise, etc.), motivating a simulator that can capture both consistently.

This paper documents the simulator setup and the hardware models used in the current 6GTandem modeling framework \cite{6gtandemD31,6gtandemD32,6gtandemD34}.
The main contributions are: i) A modular, configuration-driven simulation workflow for sub-\gls{thz} radio stripes and ii) Hardware-aligned component models parameterized by measurement/simulation data (S-parameters, gain/noise figures, nonlinear transfer characteristics).

\subsection{Related Work}
End-to-end evaluation of next-generation wireless systems is supported by network simulators, channel simulators, and link-level frameworks, each with different abstraction levels.
For example, the ns-3 mmWave module enables full-stack simulation of cellular mmWave networks with modular \gls{phy}/\gls{mac} components and statistical channels, with optional support for measurement- or ray-tracing-based traces \cite{8344116}.
NYUSIM provides an open-source, measurement-driven mmWave and sub-\gls{thz} channel simulator \cite{10367974}, and has been integrated into ns-3 to enable drop-based sub-\gls{thz} end-to-end simulations (e.g., at \SI{140}{\giga\hertz}) for throughput/latency studies using an existing protocol stack \cite{10570665}.
Complementary \gls{phy}-layer frameworks such as HermesPy support link-level evaluation and can include hardware-in-the-loop experiments \cite{9950269}, while impairment-focused studies investigate how sub-\gls{thz} phase noise shapes waveform and receiver design \cite{10604898}.
Outside communications, modular end-to-end simulators in submillimeter domains (e.g., astronomical instrumentation) also demonstrate the value of system models that span multiple components \cite{huijten2020tiempo}.

These approaches are well suited for their respective goals, but they do not capture the specific coupling introduced by the \emph{radio-stripe} architecture, where distributed \glspl{ru} are connected by a fiber-based analog transport chain and where algorithmic choices (e.g., \gls{ru} selection and beamforming) are directly shaped by hardware effects (e.g., frequency-selective fiber transport, insertion-loss accumulation, nonlinear amplification, and phase noise).
In contrast, this simulator combines (i) a configurable waveform layer, (ii) measurement-aligned component models along the fiber/stripe chain, and (iii) either stochastic channel models or site-specific ray-tracing \gls{csi} datasets, enabling consistent studies of the interplay between algorithms and hardware-limited signal propagation. Furthermore, the simulator allows to include sub-\SI{10}{\giga\hertz} access points to investigate trade-offs and complementary dual-band operation for communication and positioning.

\section{Simulator Overview}
\subsection{Scope and architecture}
The simulator implements an end-to-end signal chain and is organized around four layers:
\begin{enumerate}
    \item \textbf{Scenario definition}: environment geometry, stripe and \gls{ue} locations, and carrier/bandwidth settings are described in YAML configuration files.
    \item \textbf{Waveform layer}: the default waveform is \gls{cp}-\gls{ofdm}, including pilot insertion and oversampling.
    \item \textbf{Hardware chain}: baseband \gls{iq} signals propagate from the \gls{cu}, over the fibers and through \glspl{ru}/\glspl{bu} using configurable component models.
    \item \textbf{Wireless propagation}: sub-\gls{thz} channels can be loaded from precomputed datasets (e.g., ray-tracing) or generated using stochastic or deterministic models (e.g., Rayleigh, \gls{los}).
\end{enumerate}

\subsection{Simulation Setup Flow}
The overall workflow consists of configuring the simulator through YAML files (\cref{sec:config-file}), generating the channel response and performing the waveform/hardware simulation (\cref{sec:python-simulator}). Either a \gls{rt} (\cref{sec:3d-models-rt,sec:sionnaRT}) computes the channels upfront to be used later in the hardware simulation or deterministic/stochastic models (\cref{sec:channel-mdoels}) compute the channel on the fly while simulating the hardware.
Sionna \gls{rt} generates the channel response upfront and exports it as a dataset; the Python simulator then applies the channels by lookup while running the waveform and hardware models (\cref{sec:wireless-channel,sec:hardware-models}).

\subsubsection{Configuration File}\label{sec:config-file}
Simulation scenarios are configured through human-readable YAML files shipped with the simulator repository.
The configuration is split into (i) an environment file defining geometry and node locations, (ii) waveform configuration file, and (iii) component configuration file, allowing re-use of waveform and hardware models across multiple rooms/stripe layouts.
Currently, parameters related to the antennas, carrier frequency, bandwidth and number of subcarriers are tied to the \gls{rt} configuration. Coupling the \gls{rt} more tightly to the simulator, allowing it to run in real time, would overcome this limitation. For the time being the same parameters have to be used in the simulator as in the specific environment simulation in the `RT\_module`.

\paragraph{Environment configuration}
Each environment folder contains a configuration file with the following main sections:
\begin{itemize}
    \item \texttt{room}: room dimensions in meters (\texttt{x}, \texttt{y}, \texttt{z}).
    \item \texttt{stripe\_config}: high-level stripe topology (number of stripes \texttt{N\_stripes}, number of \glspl{ru} \texttt{N\_RUs}, inter-\gls{ru} and inter-stripe spacing, start/end positions and orientation metadata).
    \item \texttt{radio\_stripes}: a list of stripes, where each stripe is a list of nodes starting with one \gls{cu} (\texttt{central\_unit}) followed by the \glspl{ru} (\texttt{radio\_unit}), each specified by $(x,y,z)$ coordinates.
    \item \texttt{ue\_positions}: a list of \gls{ue} locations, each specified by $(x,y,z)$ coordinates.
    \item \texttt{sub\_thz}: optional frequency-band metadata for the sub-\gls{thz} mode used by the \gls{rt}. This includes carrier frequency \texttt{fc} ($f_\mathrm{c}$), bandwidth \texttt{bw} ($B$), and number of subcarriers \texttt{num\_subcarriers} ($Q$).
    \item \texttt{sub10GHz}: optional configuration for a sub-\SI{10}{\giga\hertz} baseline (e.g., \glspl{ap} positions and antenna pattern/polarization) and the number of antenna branches per \gls{ap} (\texttt{N\_sub10ghz\_antennas}). Carrier frequency \texttt{fc} ($f_\mathrm{c}$), bandwidth \texttt{bw} ($B$), and number of subcarriers \texttt{num\_subcarriers} ($Q$) are only added when a \gls{rt} generated the channel.
    \item \texttt{antenna}: number of antenna branches per \gls{ru} (\texttt{N\_antennas}), antenna polarization (\texttt{polarization}) and pattern selection (\texttt{pattern}). Only present when \gls{rt} is used.
    \item \texttt{central\_unit\_fiber\_length}: fiber length between the \gls{cu} and the first \gls{ru}.
\end{itemize}

\paragraph{Waveform configuration}
The waveform file defines the baseband numerology, e.g., \texttt{waveform\_type} (e.g., \gls{cp}-\gls{ofdm}), \texttt{n\_ofdm\_symbols}, \texttt{qam\_order}, \texttt{oversampling\_factor}, \texttt{cp\_length}, pilot settings (\texttt{pilot\_spacing}, \texttt{pilot\_mode}), and transmit power (\texttt{tx\_power}).

\paragraph{Component configuration}
Hardware blocks are parameterized through a component YAML file that selects models and parameters for the stripe chain, e.g., booster and antenna-side amplifier settings (\texttt{boost\_amplifier}, \texttt{antenna\_amplifier}), fiber (\texttt{fiber}) and coupler (\texttt{coupler}) models, and their corresponding measurement files (e.g., \texttt{.s2p} responses).

\subsubsection{3D Models and Ray Tracer}\label{sec:3d-models-rt}
Geometry-based propagation is handled by a companion module (``RT\_module'') that runs offline.
The ray tracer computes multipath components and exports a per-subcarrier \gls{mimo} channel to a NetCDF dataset. A 3D scene description specifies the simulation environment in an XML format and assigns the correct material properties. The environment YAML defines the stripe and \gls{ue} locations for which to compute the channels.
During simulation, the Python layer reuses the same configuration to query $\mat{H}[q]$ for a given stripe/\gls{ru} and \gls{ue} index, keeping the computationally heavy ray tracing step out of the main loop.
Implementation details of the Sionna \gls{rt} setup and dataset structure are given in \cref{sec:sionnaRT}.

\subsubsection{System Simulator}\label{sec:python-simulator}

\Cref{fig:Tandem_concepts} shows the modeled signal flow along a stripe.
Only one \gls{ru} is assumed to be active at any given time during transmission, while preceding units act as \glspl{bu} to compensate insertion losses and fiber attenuation.
The simulator supports gain calibration routines that set booster gains to maintain a desired power level along the stripe.
The Python simulator integrates the scenario configuration, waveform generation, hardware-aware component models, and wireless propagation into a single end-to-end execution flow:
\begin{enumerate}
    \item \textbf{Waveform generation:} generate baseband symbols and the time-domain \gls{cp}-\gls{ofdm} waveform (including pilots and oversampling) from the waveform configuration.
    \item \textbf{Stripe and hardware chain:} instantiate the stripe from the configured node locations and apply the selected component models (fiber/couplers, booster and antenna-side amplifiers, splitter/combiner, phase shifters) as the signal propagates from the \gls{cu} via the \glspl{bu} to the active \gls{ru}.
    \item \textbf{Wireless channel integration:} apply the configured wireless channel model (stochastic/deterministic) or load precomputed \gls{rt} channels; for \gls{ofdm}, the channel is applied per subcarrier via $\vect{y}[q]=\mat{H}[q]\vect{x}[q]$ (see \cref{sec:sionnaRT}).
\end{enumerate}
The detailed model options for the hardware chain and the wireless channel are described in \Cref{tab:models} and \cref{sec:sionnaRT}, respectively. A complete overview of how to implement a simulation can be found in the examples folder of the simulator repository. For example, the \emph{examples/stripe\_ul/example.ipynb} notebook shows how to configure and run an uplink simulation and plots intermediate results as the signal travels along the stripe.

\section{Hardware-Aware Component Models}\label{sec:hardware-models}
\subsection{Model selection and parameterization}
All hardware blocks inherit from a common \texttt{Component} abstraction that supports multiple modeling modes (e.g., \texttt{ideal}, odd-order polynomial nonlinearity, soft limiting, etc.).
Scenario-level YAML files select component models and parameters; for example, S-parameter files (\texttt{.s2p}) to define frequency-selective couplers and \gls{pmf} links, while amplifier modes and noise figures are configured independently. This allows for easy integration of new measurements or model updates into the simulator framework.

\begin{table*}[t]
\centering
\caption{Implemented component models and typical parameter sources.}
\label{tab:models}
\begin{tabular}{@{}p{0.15\textwidth}p{0.55\textwidth}p{0.10\textwidth}@{}}
\toprule
\textbf{Block} & \textbf{Model options} & \textbf{Reference} \\ %
\midrule
Fiber/\gls{pmf} & Fixed damping, linear filter, time/frequency-domain application, propagation delay & \cref{sec:fiber-coupler} \\ %
Coupler & Fixed damping, linear filter, time/frequency-domain application & \cref{sec:fiber-coupler} \\ %
Amplifier (PA) & \texttt{ideal}, \texttt{tanh}/\texttt{atan}, odd-order polynomial, soft limiter + noise figure & \cref{sec:amplifier} \\ %
DAC & Clipping + uniform quantization (\#bits) & \cref{sec:central-unit} \\ %
Oscillator & \texttt{ideal}, \texttt{cfo}, phase-noise models (AR/spectrum) & \cref{sec:central-unit} \\ %
IQ modem & Static/filter-based \gls{iq} imbalance + DC offset & \cref{sec:central-unit} \\ %
Splitter/combiner & Power splitting ($1/\sqrt{N}$), coherent combining & \cref{sec:ru-bu} \\ %
Phase shifter & Per-element complex phase, steering-vector phases & \cref{sec:ru-bu} \\ %
Antenna Pattern & Measured pattern, 3GPP TR 38.901 model & \cref{sec:antenna} \\ %
\bottomrule
\end{tabular}
\end{table*}

\subsection{Central Unit}\label{sec:central-unit}
The \gls{cu} generates and processes complex baseband samples prior to injection into the first fiber segment.
In the current implementation, the main impairments are:
(i) DAC clipping/quantization,
(ii) \gls{iq} imbalance and DC offset,
(iii) oscillator phase noise and \gls{cfo}, and
(iv) amplifier nonlinearity and noise.
In compact form, the modeled \gls{cu} chain can be written as
\begin{equation}
    \vect{y} = Gf_{\mathrm{PA}}\!\left(\mathcal{M}_{\mathrm{IQ}}\!\left(\mathcal{Q}\!\left(\vect{x}\right), e^{j\boldsymbol{\phi}}\right)\right),
\end{equation}
where $\mathbf{x}$ is a vector of baseband samples, $\mathcal{Q}(\cdot)$ models DAC effects, $\mathcal{M}_{\mathrm{IQ}}(\cdot)$ applies \gls{iq} impairment and mixing by the oscillator phasor $e^{j\boldsymbol{\phi}}$, and $f_{\mathrm{PA}}(\cdot)$ is the transmit power amplifier model and $G$ represents the linear gain.

\subsection{Fiber and couplers}\label{sec:fiber-coupler}
Fiber segments and couplers are modeled as linear filters with frequency selectivity, gain, and delay \cite{6gtandemD31,6gtandemD32}. When S-parameters are available, the simulator interpolates the measured/simulated $S_{21}(f)$ to the \gls{ofdm} subcarrier grid and applies it either:
(i) in the frequency domain as multiplication per subcarrier, or
(ii) in the time domain through convolution with the corresponding impulse response.
The frequency domain implementation can be written as
\begin{align}
    \vect{y} = \vect{h}^{(f)} \odot \vect{x}
\end{align}
where $\vect{h}^{(f)} \in \mathbb{C}^{Q\times 1}$ represents the fiber or couplers frequency response at the $Q$ sub-carriers and $\odot$ represents the Hadamard or element-wise product. When considering the time-domain implementation, it can be written as
\begin{align}
\vect{y} = \vect{h}^{(t)} \ast \vect{x}
\end{align}
where $\vect{h}^{(t)} \in \mathbb{C}^{L\times 1}$ is the discrete-time impulse response of the fiber or coupler that is obtained by applying the \gls{ifft} to  $\vect{h}^{(f)}$.  

\begin{figure}[h]
\centering
\begin{tikzpicture}
\begin{axis}[
    width=\columnwidth,
    height=0.4\columnwidth,
    xlabel={Frequency [GHz]},
    ylabel={Amplitude [dB]},
    legend style={at={(0.9,0.1)}, anchor=south east},
    legend cell align={left},
    legend columns=2,
]
\addplot[color1] table [x expr=\thisrow{frequency}/1e9, y expr=10*log10(\thisrow{magnitude}), col sep=comma, each nth point=100] {models/pmf1m.csv};
\addplot[color2] table [x expr=\thisrow{frequency}/1e9, y expr=10*log10(\thisrow{magnitude}), col sep=comma, each nth point=100] {models/pmf2m.csv};
\legend{PMF 1 m, PMF 2 m}
\end{axis}
\end{tikzpicture}
\caption{\gls{pmf} amplitude response.}
\label{fig:pmf-amplitude}
\end{figure}
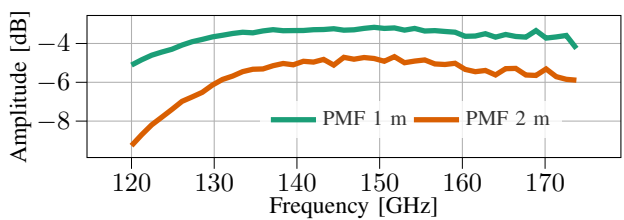

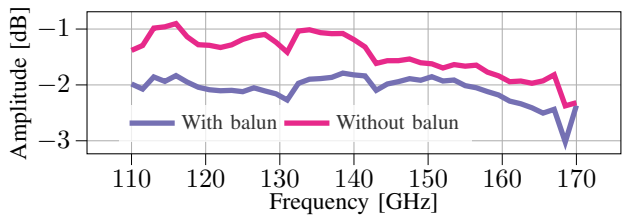
\begin{figure}[h]
\centering
\begin{tikzpicture}
\begin{axis}[
    width=\columnwidth,
    height=0.4\columnwidth,
    xlabel={Frequency [GHz]},
    ylabel={Amplitude [dB]},
   legend style={at={(0.05,0.05)}, anchor=south west},
   legend columns=2,
    legend cell align={left},
]
\addplot[color3] table [x expr=\thisrow{frequency}/1e9, y expr=10*log10(\thisrow{magnitude}), col sep=comma, each nth point=10] {models/with_balun.csv};
\addplot[color4] table [x expr=\thisrow{frequency}/1e9, y expr=10*log10(\thisrow{magnitude}), col sep=comma, each nth point=10] {models/without_balun.csv};
\legend{With balun, Without balun}
\end{axis}
\end{tikzpicture}
\caption{Coupler amplitude response.}
\label{fig:coupler-amplitude}
\end{figure}

The propagation delay is derived from the fiber length and sampling rate, enabling consistent timing across the stripe.
In the current hardware development, coupler insertion loss is a key sensitivity parameter (reported around \SI{7}{\decibel} in early designs, with a target of \SI{4}{\decibel}--\SI{5}{\decibel}) \cite{6gtandemD32}, and is therefore explicitly configurable.
Representative \gls{pmf} and coupler amplitude responses used for parameterization are shown in \Cref{fig:pmf-amplitude,fig:coupler-amplitude}.

\subsection{Radio Units with boost and active mode}\label{sec:ru-bu}
As illustrated in~\cref{fig:Tandem_concepts}, each \gls{ru} contains two logical modes: \textbf{booster mode} (forwarding along the fiber) and \textbf{active mode} (feeding the antenna array). In booster mode, the chain is: input coupler $\rightarrow$ booster amplifier $\rightarrow$ output coupler. In active mode, the chain is: input coupler $\rightarrow$ splitter $\rightarrow$ per-element phase shifters $\rightarrow$ antenna-side amplifier. The splitter applies $1/\sqrt{N}$ attenuation for $N$ antenna branches, while the combiner coherently sums received branches. The current phase shifter implementation supports arbitrary phase shifts. However, due to the component-wise implementation, phase-shifter behavior (insertion loss, tuning range, and bandwidth) can be updated in future models to reflect the observed differences between passive and active phase-shifter designs in D-band prototypes \cite{6gtandemD32}.

\subsection{Amplifier models}\label{sec:amplifier}
\begin{figure}[h]
\centering
\begin{tikzpicture}
\begin{axis}[
    width=\columnwidth,
    height=0.6\columnwidth,
    xlabel={Input amplitude $|x|$},
    ylabel={Output amplitude $|y|$},
    legend columns=3,
    legend style={at={(0.8,0.2)}, anchor=east},
    legend cell align={left},
]
\addplot[color1, only marks, mark=*, mark size=0.4pt, each nth point=6] table [x=x, y=y, col sep=comma] {models/amplifier-2x-nf10db.csv};
\addplot[color2, only marks, mark=*, mark size=0.4pt, each nth point=6] table [x=x, y=y, col sep=comma] {models/amplifier-4x-nf10db.csv};
\addplot[color3, only marks, mark=*, mark size=0.4pt, each nth point=6] table [x=x, y=y, col sep=comma] {models/amplifier-6x-nf10db.csv};
\addplot[color4, only marks, mark=*, mark size=0.4pt, each nth point=6] table [x=x, y=y, col sep=comma] {models/amplifier-8x-nf10db.csv};
\addplot[color5, only marks, mark=*, mark size=0.4pt, each nth point=6] table [x=x, y=y, col sep=comma] {models/amplifier-10x-nf10db.csv};
\legend{$2\times$,$4\times$,$6\times$,$8\times$,$10\times$}
\end{axis}
\end{tikzpicture}
\caption{\Gls{pa} AM/AM behaviour for different gains.}\label{fig:pa-amam}
\end{figure}
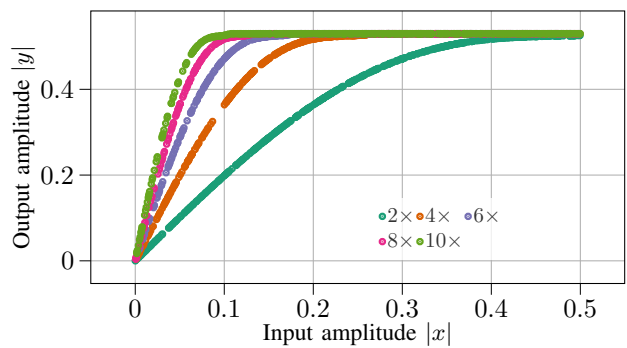

Power amplifiers are implemented as a memoryless nonlinearity with additive complex Gaussian noise, followed by a linear gain.
This is written as
\begin{equation}
    y = f(x) = G\, f_{\mathrm{PA}}(x + w),
\end{equation}
where $G$ is the small-signal gain, $w$ is \gls{awgn} derived from the configured noise figure and bandwidth, and $f_{\mathrm{PA}}(\cdot)$ is selected from a set of nonlinear functions (e.g., odd-order polynomial, soft limiter, etc.). For polynomial modes, coefficients can be fitted to measured transfer characteristics (e.g., D-band \gls{pa} characterization around \SI{150}{\giga\hertz}) \cite{6gtandemD32}. As an example of a polynomial fit to a developed \gls{pa}, the \gls{amam} characteristics of the full \gls{pa} model (including the small-signal gain)  for a varied gain are shown in \Cref{fig:pa-amam}. Future work can incorporate memory-based models such as Volterra series or the generalized memory polynomial.

\subsection{Antenna Model}\label{sec:antenna}
Sub-THz antenna patterns are directly incorporated in the ray-tracer wireless channel datasets. For the wireless channel generation, Sionna RT is used as discussed in~\cref{sec:sionnaRT}. Antenna patterns can be incorporated from measured or simulated data. Note that one restriction of the currently used raytracer is the fact that antenna patterns cannot be applied on a per-antenna basis, as all antenna elements require the same pattern in Sionna RT. Nevertheless, the channel is comuted between each transmit and receive antenna pair, allowing steering by applying phase shifts within our simulator. Currently, two antenna patterns are supported, namely the built-in Sionna TR 38.901 model and antenna pattern stemming from developed antennas within the 6GTandem project~\cite{10501515}.

\section{Wireless Channel Integration}\label{sec:wireless-channel}

Two integrations are supported in the simulator: i) channel models and ii) spatially consistent \gls{rt} channels. The latter is precomputed and loaded into the simulator at runtime.

\subsection{Channel Models}\label{sec:channel-mdoels}

The simulator provides lightweight, synthetic wireless channel models that can be used when no site-specific \gls{csi} is available. They do not fully capture all real-world behaviors but their value lies in the short simulation times for quick simulations and debugging.
All models operate in the \gls{ofdm} frequency domain and produce a per-subcarrier \gls{mimo} channel matrix $\mat{H}[q]$.

For each subcarrier index $q$ with (absolute) frequency $f_q$ and for each \gls{tx}/\gls{rx} antenna pair $(m,k)$, the frequency-domain channel coefficient is decomposed into
(i) a large-scale free-space spreading gain, (ii) optional direction-dependent antenna gains, and (iii) a small-scale fading term.
The free-space (power) path gain is modeled as
\begin{align}
    \beta_{\mathrm{fs}}(d_{m,k},f_q)
    &= \left(\frac{c}{4\pi f_q d_{m,k}}\right)^2.
\end{align}

The overall channel coefficient is then given by
\begin{align}
    h_{q, m,k}
    = g_{q, m,k}\sqrt{G^{\mathrm{tx}}_{m}\,G^{\mathrm{rx}}_{k}\,\beta_{\mathrm{fs}}(d_{m,k},f_q)},
\end{align}
where $d_{m,k}$ is the distance between the antenna phase centers and $c$ is the speed of light.
The antenna gains $G^{\mathrm{tx}}_{m}$ and $G^{\mathrm{rx}}_{k}$ are optional (set to $1$ for isotropic antennas).
The small-scale fading term $g_{q,m,k}$ captures the remaining propagation effects and is normalized such that $\expt{\lvert g_{q,m,k}\rvert^2}=1$.
The per-subcarrier channel matrix is constructed as $\left[\mat{H}[q]\right]_{k,m}=h_{q,m,k}$ and applied as $\vect{y}[q]=\mat{H}[q]\vect{x}[q]$.

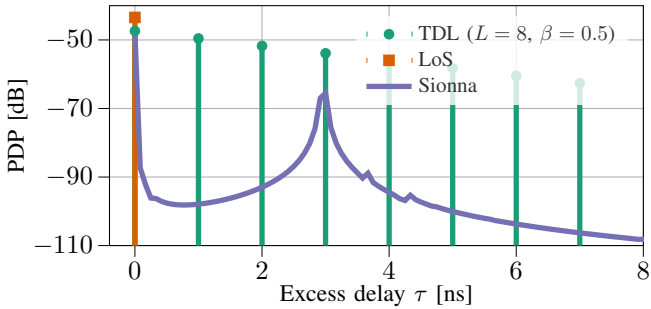
\begin{figure}[t]
\centering
\begin{tikzpicture}
\begin{axis}[
    width=\columnwidth,
    height=0.55\columnwidth,
    xlabel={Excess delay $\tau$ [ns]},
    ylabel={PDP [dB]},
    xmin=-0.4, xmax=8,
    ymin=0, ymax=70,
    y filter/.expression={y+110},
    yticklabel={\pgfmathparse{\tick-110}\pgfmathprintnumber{\pgfmathresult}},
    grid=both,
    legend cell align={left},
    legend style={at={(0.98,0.98)}, anchor=north east},
]
\addplot[color1, ycomb, mark=*, mark size=1.0pt] coordinates {
    (0, -47.4) (1, -49.57) (2, -51.74) (3, -53.91) (4, -56.09) (5, -58.26) (6, -60.43) (7, -62.6)
};
\addlegendentry{TDL ($L=8$, $\beta=0.5$)}
\addplot[color2, ycomb, mark=square*, mark size=1.2pt] coordinates {(0, -43.43)};
\addlegendentry{LoS}
\IfFileExists{figures/pdp-0.73x--0.71y-2.00z.csv}{
  \addplot[color3] table [col sep=comma, x expr=\thisrowno{1}*1e9, y expr=\thisrowno{2}] {figures/pdp-0.73x--0.71y-2.00z.csv};
  \addlegendentry{Sionna}
}{}
\end{axis}
\end{tikzpicture}
\caption{Example power-delay profiles for the exponential \gls{tdl} model and the \gls{los} model at $f_\mathrm{c}=\SI{157.75}{\giga\hertz}$ and $d=\SI{2.24}{\meter}$ (free-space spreading, isotropic antennas). Antenna gains $G_\mathrm{tx}=\SI{20.00}{\decibel}$ and $G_\mathrm{rx}=\SI{20.00}{\decibel}$ are applied. For the \gls{tdl}, the excess delays are $\tau_l = l\Delta\tau$ with tap spacing $\Delta\tau=\SI{1}{\nano\second}.$}
\label{fig:pdp-los-tdl}
\end{figure}

\Cref{fig:pdp-los-tdl} shows example power-delay profiles for the exponential \gls{tdl} model and the \gls{los} baseline.

\subsubsection{Deterministic Channel Models}
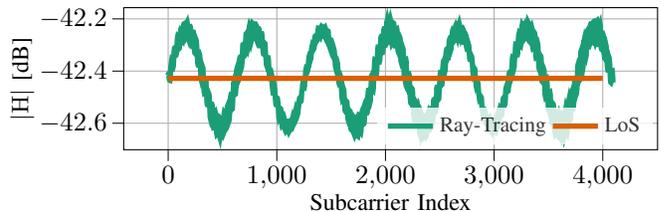
\begin{figure}
    \centering
    \begin{tikzpicture}
        \begin{axis}[
            width=\columnwidth,
            height=0.4\columnwidth,
            xlabel={Subcarrier Index},
            ylabel={$|\mathrm{H}|$ [dB]},
            legend style={at={(1.0,0)}, anchor=south east},
            legend cell align={left},
            legend columns=2,
        ]
        \addplot[color1] table [x=subcarriers, y=h_sionna, col sep=comma] {figures/los-channel-comparison.csv};
        \addplot[color2] table [x=subcarriers, y=h_los, col sep=comma, each nth point=100] {figures/los-channel-comparison.csv};
        \legend{Ray-Tracing, \gls{los}}
        \end{axis}
    \end{tikzpicture}
    \caption{The comparison of the \gls{los} with \gls{rt} channels prove that they are accurate even at sub-THz frequencies.}
    \label{fig:los-model}
\end{figure}

A lightweight deterministic \gls{los} baseline is included that is useful for initial experiments and validation of algorithms.
In the \gls{los} model, the small-scale term is a unit-magnitude phase rotation,
\begin{align}
    g^{\mathrm{LoS}}_{q,m,k}
    &= 
    \exp\!\left(-j\frac{2\pi f_q}{c} d_{m,k}\right).
\end{align}
This yields $h_{q,m,k}=\sqrt{G^{\mathrm{tx}}_{m}G^{\mathrm{rx}}_{k}\beta_{\mathrm{fs}}(d_{m,k},f_q)}\,g^{\mathrm{LoS}}_{q,m,k}$.
For narrowband approximations, one can set $f_q \approx f_\mathrm{c}$ for all subcarriers to obtain a frequency-flat \gls{los} channel. A comparison with the \gls{rt} channels (\cref{sec:sionnaRT}) demonstrate that this model holds at sub-THz frequencies (\cref{fig:los-model}). The antenna gains $G^{tx}_{m}, G^{rx}_{k}$ are calculated based on the angle between the \gls{ue} and \gls{ru}.

\subsubsection{Stochastic Channel Models}
Currently, two stochastic small-scale fading models are supported. 

\paragraph{Uncorrelated Rayleigh Fading}
First, an uncorrelated Rayleigh fading model is considered, where the small-scale coefficients are drawn independently across $(q,m,k)$ as
\begin{align}
    g_{q,m,k} \sim \mathcal{CN}(0, 1).
\end{align}

\paragraph{Rayleigh \gls{tdl} Channel}
Next to this, a Rayleigh \gls{tdl} channel is considered. The time domain channel impulse response between \gls{tx} antenna $m$ and \gls{rx} antenna $k$ is modeled as 
 \begin{align}
     \vect{g}_{m,k}^{(t)}& = \mat{C_h}^{1/2} \vect{\tilde{g}}_{m, k},\\
     \vect{\tilde{g}}_{m,k} &\sim \mathcal{CN}(\vect{0}, \vect{I}_{L}).
 \end{align}
Here $\vect{\tilde{g}}_{m,k}^{(t)} \in \mathbb{C}^{L \times 1}$ consists of $L$ uncorrelated taps with an exponentially decaying \gls{pdp}. The tap powers are
\begin{align}
    \mat{C_h} &= \diag{\lambda_h^0, \cdots, \lambda_h^{L-1}},\\
    \lambda_h^l &= e^{-\beta l},
\end{align}
where $\beta$ controls the amount of frequency selectivity (high $\beta$ yields flatter channels).
The \gls{pdp} of tap $l$ is $\expt{\left|g_l^{(t)}\right|^2}=\lambda_h^l$, and we normalize the overall tap power by setting $\tr{\mat{C_{h}}}=1$.
Assuming the cyclic prefix is longer than the maximum excess delay, the frequency response on a $Q$-subcarrier grid is obtained by a $Q$-point \gls{dft},
\begin{align}
    \vect{g}^{(f)}_{m,k} = \mat{F}\mat{\Sigma}\vect{g}_{m,k}^{(t)},
\end{align}
where $\mat{F}\in \mathbb{C}^{Q \times Q}$ is the \gls{dft} matrix and
\begin{align}
    \mat{\Sigma}=
    \left[\begin{array}{c}
        \mathbf{I}_L \\
        \mathbf{0}_{(Q-L) \times L}
    \end{array}\right]
\end{align}
zero-pads $\vect{g}^{(t)}_{m,k}$ to length $Q$.
The per-subcarrier small-scale fading coefficient is then $g_{q,m,k}=\left[\vect{g}^{(f)}_{m,k}\right]_q$.

\subsection{Ray-Tracing Channels using Sionna}\label{sec:sionnaRT}
To complement stochastic fading, the simulator can load deterministic, site-specific \gls{mimo} channel state information generated with the NVIDIA Sionna \gls{rt} engine~\cite{sionna-rt-tech-report, sionna}.
In the simulator, this functionality is provided as a companion module (``RT\_module'') that precomputes channels for a given 3D scene and exports them in a dataset format that the simulator can query efficiently at runtime.
The overall goal is to make the wireless propagation model consistent with (i) the room geometry, (ii) the radio-stripe topology, and (iii) the \gls{ofdm} numerology used by the waveform layer, while keeping the computationally heavy ray-tracing step out of the main simulation loop. 

\paragraph{Scene import and material modeling}
The RT pipeline loads an indoor environment from a scene description (XML) and assigns radio materials to the scene objects.
To capture strong frequency dependence at sub-\gls{thz}, the material parameters (conductivity and relative permittivity) are updated as a function of carrier frequency using ITU-inspired callbacks for common building materials (e.g., glass, concrete, metal, polystyrene, and MDF) as measured in~\cite{Hexa_D23}. This enables the same geometry to be reused for sub-\SI{10}{\giga\hertz} and sub-\gls{thz} operation with frequency-appropriate electromagnetic parameters. 
\paragraph{Transmitter/receiver placement and antenna arrays}
The radio-stripe layout is taken from a stripe configuration (number of stripes, number of \glspl{ru} per stripe, inter-\gls{ru} and inter-stripe spacing, and a stripe start position). As the component placement determines the channel and stripe characteristics this configuration is shared between the simulator and the RT setup.
For each \gls{ru} position, the RT pipeline instantiates a Sionna \texttt{Transmitter} with a planar antenna array; the \gls{ue} is instantiated as a \texttt{Receiver} at a configurable set of locations.
In the default dataset generation, both ends use a $1\times4$ uniform linear array and a single polarization, which yields a per-subcarrier channel matrix with dimension $(N_\mathrm{rx})\times(N_\mathrm{tx})$, where $N_\mathrm{rx}=N_\mathrm{tx}=4$. A $1\times4$ array matches the hardware designed in the 6GTandem project D4.3 \cite{6gtandemD43}. However, this can be adapted to any desired array configuration.

\paragraph{Ray-tracing and frequency response generation}
For each \gls{ue} location and each \gls{ru}, the RT pipeline runs Sionna's \texttt{PathSolver} with line-of-sight enabled and a finite interaction depth (default maximum depth 5), including specular reflections and refraction while disabling diffuse scattering. 
The resulting set of multipath components is converted to a \gls{cfr} on the subcarrier grid.
The subcarrier frequencies are aligned with the simulator waveform settings; for the current sub-\gls{thz} configuration this corresponds to $f_\mathrm{c}=\SI{157.75}{\giga\hertz}$, $B=\SI{3.0}{\giga\hertz}$ and $4096$ subcarriers, i.e., the RT output can be applied directly as $\mathbf{y}[q]=\mathbf{H}[q]\mathbf{x}[q]$ without additional interpolation.

\paragraph{Dataset structure and runtime integration}
To support reproducible experiments and fast random access during simulation, the RT output is stored as NetCDF files.
\Gls{ue} metadata is stored in a separate NetCDF file, allowing for lightweight querying of users based on their characteristics. This metadata contains the location coordinates, a user id and optional metadata indices for locations directly under a stripe grid.
For each \gls{ue} index, a channel file stores the complex \gls{cfr} tensor together with coordinates that map each transmitter instance to its stripe and \gls{ru} id.
At runtime, the simulator matches the requested \gls{ue} coordinates to the \gls{ue} index, loads the corresponding channel file, and extracts $\mathbf{H}[q]$ for a given stripe/\gls{ru} pair on demand.

\section{Configuration and Simulation Output}
The simulation is driven by a set of YAML configuration files (\cref{sec:config-file}).
Separating scenario, waveform, and component parameters enables consistent experiments across environments and straightforward parameter sweeps without modifying the simulator code.

This supports studying the impact of hardware and waveform choice in a specific scenario. Only certain parts of a stripe can be enabled or bypassed to analyze specific effects. Instantiating components with an ideal/linear model forms a reference to compare the hardware-accurate models against. These realistic models are parametrized by measurements e.g., S-parameters, gain/noise figures, and nonlinear transfer characteristics.
The wireless side supports switching between deterministic/stochastic channel models and ray-tracing datasets; pilot density and mode (e.g., scattered pilots within each \gls{ofdm} symbol versus dedicated pilot-only \gls{ofdm} symbols spanning all subcarriers) determine the quality of channel estimates, and enable controlled studies of channel estimation error.
Finally, the same workflow can be repeated for different scenarios changing the environment, stripe layout, beamforming strategy, and \gls{ru} selection policy to quantify geometry- and activation-dependent performance.

\begin{figure}[h]
    \centering
    \begin{tikzpicture}
        \begin{axis}[
            width=\columnwidth,
            height=0.4\columnwidth, %
            colorbar horizontal,
            enlargelimits=false,
            xlabel={Stripe index},
            ylabel={\gls{ru} index},
            colormap/viridis,
            colorbar style={
                xlabel={NMSE [dB]}
            }
        ]
        \addplot[
            matrix plot,
            point meta=explicit,
            mesh/cols=20,
        ] table [x=ru_id, y=stripe_id, meta=sndr_cu, col sep=comma] {figures/heatmap.csv};
        \draw[
            thick,
            red
        ] (axis cs:10.5,-0.5) rectangle (axis cs:11.5,0.5);
        \end{axis}
    \end{tikzpicture}
    \caption{Heatmap showing the \gls{nmse} at the \gls{cu} for each individual active \gls{ru}. Note that this is not the same as \gls{nmse} at the \gls{ru}. The \gls{ue} position is indicated in red.}\label{fig:heatmap}
\end{figure}
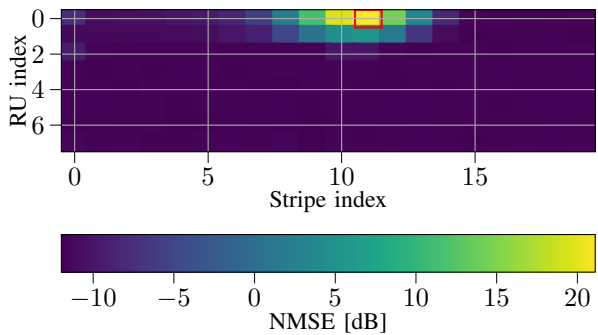

To support debugging and quantitative evaluation, the simulator produces both intermediate and end-to-end outputs.
It can export the complex baseband waveform at the input and output of each modeled component, from which constellation diagrams, AM/AM, and AM/PM characteristics can be derived.
Additionally it reports performance metrics at the \gls{cu} such as \gls{nmse} and \gls{sndr} as well as \gls{ber} after demodulation.
The frequency responses used by the component models (e.g., S-parameter traces) are stored alongside the results to facilitate easily adding new measurements as well as validation and documentation.
Simulation outputs are exported in lightweight formats (e.g., CSV) for post-processing and plotting.
As an example, \cref{fig:amam-stages} illustrates how nonlinear distortion and noise accumulate along the stripe via stage-wise AM/AM curves. \Cref{fig:heatmap} shows a heatmap of \gls{nmse} at the \gls{cu} when activating each individual \gls{ru}, which can directly support beamforming and \gls{ru} selection studies.

\begin{figure}[h]
    \centering
    \begin{tikzpicture}
        \begin{axis}[
            width=\columnwidth,
            height=0.6\columnwidth,
            xlabel={Input amplitude $|x|$},
            ylabel={Output amplitude $|y|$},
            only marks,
            every mark/.style={
              opacity=1.0,
              fill=\pgfkeysvalueof{/pgfplots/plot color}
            },
            mark size=0.2pt,
            legend style={at={(0.0,0.9)}, anchor=west},
            legend columns=5,
        ]

        \addplot[color1, mark=*, color=color1, forget plot]
            table [x=xstage0, y=ystage0, col sep=comma]
            {figures/stripe0_am_am_sorted_red.csv};

        \addplot[color2, mark=*, forget plot]
            table [x=xstage1, y=ystage1, col sep=comma]
            {figures/stripe0_am_am_sorted_red.csv};

        \addplot[color3, mark=*, forget plot]
            table [x=xstage2, y=ystage2, col sep=comma]
            {figures/stripe0_am_am_sorted_red.csv};

        \addplot[color4, mark=*, forget plot]
            table [x=xstage3, y=ystage3, col sep=comma]
            {figures/stripe0_am_am_sorted_red.csv};

        \addplot[color5, mark=*, forget plot]
            table [x=xstage4, y=ystage4, col sep=comma]
            {figures/stripe0_am_am_sorted_red.csv};

        \addlegendimage{only marks, color1, mark=*, mark options={draw=none}, mark size=0.2pt}
        \addlegendentry{Stage 1}
        \addlegendimage{only marks, color2, mark=*, mark options={draw=none}, mark size=0.2pt}
        \addlegendentry{Stage 2}
        \addlegendimage{only marks, color3, mark=*, mark options={draw=none}, mark size=0.2pt}
        \addlegendentry{Stage 3}
        \addlegendimage{only marks, color4, mark=*, mark options={draw=none}, mark size=0.2pt}
        \addlegendentry{Stage 4}
        \addlegendimage{only marks, color5, mark=*, mark options={draw=none}, mark size=0.2pt}
        \addlegendentry{Stage 5}

        \end{axis}
    \end{tikzpicture}
    \caption{AM/AM plots per stage show how PA saturation and noise affect the signal as it travels along the stripe.}
    \label{fig:amam-stages}
\end{figure}
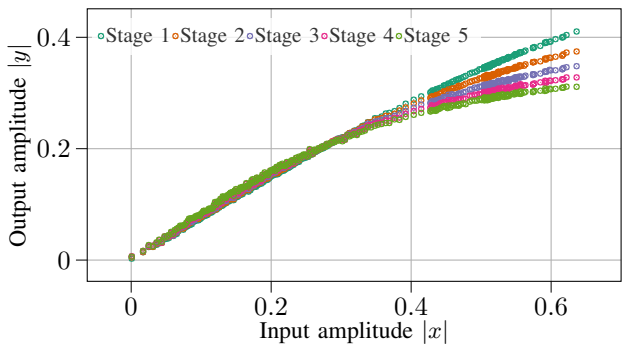

\section{Research Opportunities}

The simulator provides a waveform-level, hardware-aware testbed where system algorithms (e.g., \gls{ru} selection, beam management, localization) can be evaluated under the same cascaded front-haul and RF impairment models that ultimately shape end-to-end performance.
This enables research that connects architectural decisions to component-level constraints, while remaining reproducible via configuration files and measurement-parameterized models.

\paragraph{RU selection and beam management}
The stripe simulator makes it possible to benchmark \gls{ru} selection and beam-management policies end-to-end, from baseband through the cascaded front-haul to the wireless channel, while accounting for impairments such as nonlinearity, phase noise, and frequency-selective \gls{pmf}/coupler responses.
This is particularly relevant for approaches that use sub-\SI{10}{\giga\hertz} side information to control sub-\gls{thz} operation (e.g., learning-aided selection), where the value of side information must be quantified under realistic hardware constraints \cite{gupta2025deeplearningsubthzradio,vanderperre2025distributeddeploymentdualfrequencyconcepts}.

\paragraph{Distributed deployment and antenna/coverage co-design}
By parameterizing stripe layouts, \gls{ru} spacing, and antenna configurations, the simulator can be used to explore distributed deployment strategies that trade off coverage, spatial reuse, and hardware cost/complexity.
It supports studying how polarization, array concepts, and topology choices interact with front-haul losses and calibration requirements in dense sub-\gls{thz} deployments \cite{10501515,vanderperre2025distributeddeploymentdualfrequencyconcepts}.

\paragraph{Positioning and localization over radio-over-fiber}
The availability of a controlled, reproducible front-haul and a wireless signal chain enables systematic evaluation of distance estimation and uplink positioning algorithms for radio-over-fiber systems.
In particular, the simulator can expose such algorithms to cascaded front-haul effects (e.g., accumulated distortion and frequency selectivity), allowing sensitivity analyses and robustness studies beyond idealized models \cite{10694520,11299429}.

\paragraph{Cascaded hardware impairments and compensation}
Because the stripe inherently contains cascaded amplification stages, the simulator is well-suited to study how nonlinear distortion accumulates across boosters and how this impacts link metrics (e.g., \gls{evm}, spectral regrowth) and higher-level decisions (e.g., modulation/coding and \gls{ru} selection).
It can also serve as a platform to compare mitigation strategies such as model-based optimization for cascaded power amplifiers and over-the-air identification/training design \cite{moryakova2025modelinganalysisoptimizationcascaded,rottenberg2024optimaltrainingdesignovertheair}.

\paragraph{Fiber distribution, coupling, and packaging constraints}
The simulator's measurement-aligned modeling of \gls{pmf} and couplers enables end-to-end studies of D-band front-haul feasibility and sensitivity, including insertion-loss budgets, frequency selectivity, and the impact of packaging and coupling mechanisms.
This supports quantifying system implications of design choices such as advanced packaging for couplers and D-band \gls{pmf} links, and comparing alternative implementations under identical system assumptions \cite{10565206,liakonis2025efficientimplementationpmf,11043384}.

\paragraph{Circuit-to-system translation for D-band RF front-ends}
With a waveform-level chain, improvements at the circuit level (e.g., linearity or output power of an upconversion mixer) can be translated into system-level gains in a consistent way, accounting for wideband operation and interactions with other impairments.
This enables sensitivity studies that connect device metrics to end-to-end throughput, robustness margins, and calibration effort \cite{10981654}.

\paragraph{Waveform, pilot design, and calibration for ultra-wideband operation}
Large sub-\gls{thz} bandwidths amplify the need for robust training and calibration procedures that remain effective under nonlinearity and oscillator impairments.
The simulator enables controlled comparisons of pilot/training designs and calibration routines at waveform level, including the impact of cascaded hardware on estimator bias/variance~\cite{rottenberg2024optimaltrainingdesignovertheair}.

\section{Conclusion}
We presented an open-source, configuration-driven, hardware-aware sub-\gls{thz} radio-stripe simulator and detailed the implemented component and impairment models.
The simulator couples a configurable \gls{cp}-\gls{ofdm} waveform layer with measurement-parameterized fiber/\gls{pmf} and coupler responses and modular \gls{cu}/\gls{ru} chains, including nonlinearity, noise, \gls{iq} imbalance, and oscillator impairments.
Wireless propagation is supported via lightweight deterministic and stochastic per-subcarrier channel models and site-specific Sionna \gls{rt} datasets, while exported stage-wise waveforms and metrics (e.g., \gls{nmse}, \gls{sndr}, \gls{ber}) enable reproducible end-to-end studies of impairment accumulation, calibration, and algorithm design.
Ongoing work expands the model library and integrates updated measurements. Currently, the available models are limited, and richer front-end models will be added as new hardware gets developed and sub-THz research is expanded. While ray tracing provides the most realistic channel model it is computationally intensive and leads to large datasets. Migrating towards a real-time \gls{rt} setup would improve future usability and ease experimental variation of \gls{rt} related parameters.

\section*{Acknowledgment}
The authors thank Chesney Buyle for preparing the 3D scenes, Buon Kiong Lau and Yu-Yan Cao for providing the antenna pattern data, and Herbert Zirath, Maciej Wojnowski and Zulaicha Parastuty for providing the hardware measurement data. The authors also thank the 6GTandem team for valuable discussions and support.

\printbibliography%

\end{document}